\documentclass[smallextended]{svjour3}       
\smartqed  
\usepackage{graphicx}
\usepackage{amsmath}
\usepackage{amssymb}
\usepackage{hyperref}
\usepackage{color}
\usepackage[numbers,sort&compress]{natbib}
\begin{document}

\title{Few-neutron systems with the long-range Casimir-Polder force
\thanks{
Work supported in part by the Brazilian agency FAPESP thematic projects 
2017/05660-0 and 2019/07767-1, and INCT-FNA Proc. No. 464898/2014-5 (RH), 
and the US NSF through a grant for ITAMP at Harvard University and 
the Smithsonian Astrophysical Observatory (JFB). 
}
}


\author{R. Higa         \and
        J. F. Babb 
}


\institute{R. Higa \at
              Instituto de F\'\i sica, Universidade de S\~ao Paulo, 
              R. do Mat\~ao 1371, 05508-090, S\~ao Paulo, Brazil\\
              \email{higa@if.usp.br}           
           \and
           J. F. Babb \at
              ITAMP, Center for Astrophysics \textbar \ Harvard \& Smithsonian, 
              MS 14, 60 Garden St., Cambridge, MA 02138, USA\\
              \email{jbabb@cfa.harvard.edu}
}

\date{Received: date / Accepted: date}

\maketitle

\begin{abstract}
In this work we present results of the long-range electromagnetic 
Casimir-Polder interactions between two neutrons, a neutron and a conducting 
wall, and a neutron between two walls. As input, we use the dynamic 
dipole polarizabilities of the neutron fitted to chiral EFT results up to 
the pion production threshold and at the onset of the Delta resonance. 
Our work can be relevant to the physics of confined ultracold neutrons 
inside bottles. 
\end{abstract}

\section{Introduction}
\label{sec:intro}

The Casimir effect is a remarkable example of a phenomenon under deep 
contemplative analysis permeating through many 
different branches of physics~\cite{Spr96,RodHuiWoo15,ForHerKar16}.
It is often cited to illustrate the non-trivial concept of zero-point 
energy, or quantum fluctuations, giving rise to an observable 
force between 
two neutral objects. In its simplest version, the attractive force between 
two parallel, conducting plates is often recalled to explain the consequences 
of quantization of oscillating modes, at the heart of quantum physics, and 
the puzzling appearances of infinities that plague quantum field theories. 
Not only a necessity to explain certain quantum phenomena such 
as the behavior of specific heat of solids or the reduction of X-ray 
scattering from crystals at ultra-low temperatures~\cite{milonni-shih-92}, 
vacuum quantum fluctuations sustain the mystery of their contribution to the 
cosmological constant, 
which differs between predictions and observations by 
many orders of magnitude~\cite{milonni-shih-92}. 

The broader meaning of the Casimir effect has its origins in experiments in 
the 1940s by Overbeek at Phillips Laboratory on quartz powder in colloid 
suspension (see~\cite{milonni-shih-92} and references therein). The 
observed asymptotic behavior of the interactions disagreed with 
the van der Waals $1/r^6$ predictions and led Casimir and Polder to explain 
the mismatch in terms of retardation effects due to the finite speed of 
light. Backed by an insight from Niels Bohr, Casimir rederived and 
reinterpreted the so-called Casimir-Polder ($1/r^7$) forces in terms of 
changes in the zero-point 
energy~\cite{refId0}. It is this latter interpretation that 
excites the curiosity and interest of scientists from many distinct 
specializations in physics. 

In atomic and molecular physics, specifically, a considerable amount of work 
has been dedicated to this subject~\cite{Bab10}. 
Here, the so-called Casimir-Polder (CP) potential~\cite{CasPol48}
for the electromagnetic interactions at very large separations
describes the effects of the finite speed of light
in mutual virtual photon mediated interactions between polarizable systems~\cite{SprKel78a,Spr96}.
Feinberg and  Sucher~\cite{FeiSuc70} rederived the CP force between two 
neutral spinless particles in terms of the exchange of two virtual photons. 
The sum of all possible frequencies of the two virtual photons, obtainable 
from quantum field theory, has the same zero-point energy interpretation 
envisaged by Casimir. The Compton scattering of (virtual) photons on the 
neutral particle constitutes the sub-amplitude for the two-photon exchange 
process and carries information on the particle substructure as discussed 
in the following section. 

In the present work, we use the terminology van der Waals (vdW) potential and Casimir-Polder potential 
in the following sense---both vdW and CP potentials
are ``long-range'' electromagnetic interactions. 
Conventionally,``vdW interactions'' refer to instantaneous Coulomb interactions. Moreover
the CP potential has as its ``small separation distance'' limit the vdW potential, the CP potential is valid for
arbitrarily increasing separations (larger than some minimum separation at which ``short-range'' interactions,
such as electron exchange in, for example, atomic physics, become negligible).

At asymptotically large separations, the CP potentials
approach simple expressions involving only $\hbar$, $c$, the
individual static polarizabilities $\alpha(0)$, and an inverse power of the separation distance (e.g. 
the behavior $1/r^7$ mentioned
above with all coefficients becomes $-23\hbar c \alpha^2 (0)/(4\pi r^7)$).
Such asymptotic CP potentials are known for two neutral polarizable systems~\cite{CasPol48,FeiSuc70},
a neutral system and a charged system~\cite{BerTar76,SprKel78a,FeiSuc83}, for an atom and a perfectly conducting wall~\cite{CasPol48}, etc.
Thus, as we set forth in an earlier paper~\cite{BabbHigaHussein17},
it is reasonable, following
an \textit{ansatz} similar to that used 
by Spruch and Kelsey~\cite{SprKel78a} for atoms, to 
write down the CP potential between two neutrons, 
a neutron and a wall, or a neutron between two walls in terms of the frequency-dependent
polarizabilities $\alpha(\omega)$, where $\omega$ is the photon frequency.

Arnold~\cite{Arn73} was the first to calculate 
effects of the CP potential---using the asymptotic $1/r^7$ potential---between 
two neutrons in nucleon-nucleon scattering; however, at that time only the static, electric dipole 
polarizability data were available with nowadays outdated values. 
We extended Arnold's idea~\cite{BabbHigaHussein17,HBH17-2} to include 
dynamic electric and magnetic dipole polarizabilities with updated 
information from low-energy chiral effective field theory analysis. 
We also performed calculations of the CP-interaction between a neutron 
and a wall, and one neutron between two walls. 
In the following we 
summarize our main results and present an outlook for future studies. 

\section{Neutron dynamic dipole polarizabilities}

Electromagnetic probes have been one of the most important tools to extract 
information about the structure of hadrons. In the low ($E\lesssim 200$~MeV) 
and intermediate ($0.2\lesssim E\lesssim 1$~GeV) energy region Compton 
scattering made significant contributions to our understanding about the 
structure of the nucleon~\cite{Hagelstein:2015egb}. 
The electromagnetic field of the photon that hits the nucleon induces a 
response that can be parametrized in terms of the generalized multipole 
polarizabilities~\cite{GUIASU1979145,Hagelstein:2015egb}, the leading 
dipole ones being inputs to our neutron-neutron CP potential. 
While dynamic dipole polarizabilities of the proton have been intensively 
studied and obtained from experiments with satisfactory precision, in the 
neutron case, one has to rely on strong isospin symmetry and bound neutron 
effects for Compton scattering on the 
deuteron~\cite{PhysRevLett.90.192501,PhysRevLett.113.262506} and 
${}^3$He~\cite{Annand:2016pzl}, or on nuclear structure uncertainties on 
neutron scattering of a large $Z$ nucleus such as 
Pb~\cite{PhysRevLett.66.1015}. 

Chiral effective field theory ($\chi$EFT), the effective theory rooted in 
the chiral symmetry of the underlying quantum chromodynamics (QCD), has 
been established as a rigorous and reliable theoretical framework to 
extract information about nucleon polarizabilities in the low-energy 
regime~\cite{Griesshammer:2012we,Hagelstein:2015egb}. 
The most updated $\chi$EFT calculation of Lensky, McGovern, and 
Pascalutsa~\cite{Lensky:2015awa} takes into account recoil corrections in 
a Lorentz-covariant way, improves convergence close to the pion production 
threshold, and includes the Delta ($\Delta$) resonance explicitly. Their 
predictions 
for the neutron dynamic electric ($\alpha_n$) and magnetic ($\beta_n$) 
dipole polarizabilities for photon energies up to $\omega_\gamma=300$~MeV 
are nearly the same (within theoretical errorbars) to the proton case, 
as expected from isospin symmetry. 
In the static limit they have $\alpha_n(0)=13.7\pm 3.1$ and 
$\beta_n(0)=4.6\pm 2.7$, in units of $10^{-4}\,{\rm fm}^3$. 

\begin{figure}
\begin{tabular}{lr}
\includegraphics[width=0.5\textwidth]{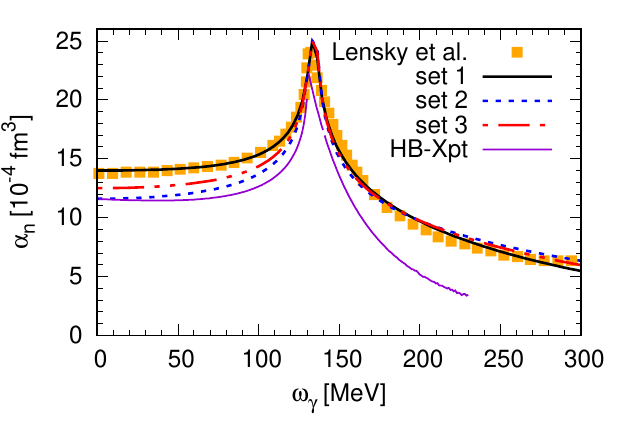}&
\includegraphics[width=0.5\textwidth]{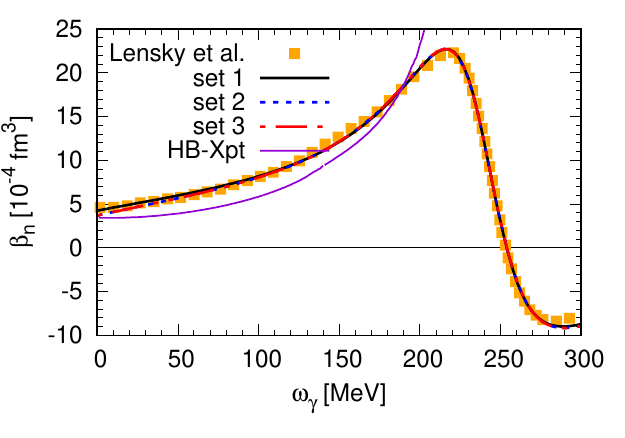}
\end{tabular}
\caption{
Dynamic electric (left) and magnetic (right) polarizabilities, as functions
of the photon energy $\omega_{\gamma}$.
The yellow squares are $\chi$EFT results of Lensky
{\em et al.}~\cite{Lensky:2015awa} while sets 1, 2, and 3 correspond to
our parametrizations using the numbers specified in Table~\ref{tab:pol-par}. 
The thin solid lines are HB-$\chi$EFT results from 
Ref.~\cite{Hildebrandt:2003fm}. 
Adapted from~\cite{BabbHigaHussein17}. 
\label{fig:n-polariz}
}
\end{figure}

The relations between the dynamic dipole polarizabilities and either Compton 
scattering observables or theory predictions are quite 
involved~\cite{Hagelstein:2015egb}. Therefore, we provide a parametrization 
of $\alpha_n(\omega)$ and $\beta_n(\omega)$ that tries to incorporate 
the relevant low-energy physics with simple formulas. 
They take the form 
\begin{eqnarray}
&&\alpha_n(\omega)=\frac{\alpha_n(0)\sqrt{(m_{\pi}\!+\!a_1)
(2M_n\!+\!a_2)}(0.2a_2)^2}{\sqrt{(\sqrt{m_{\pi}^2\!-\!\omega^2}\!+\!a_1)
(\sqrt{4M_n^2\!-\!\omega^2}\!+\!a_2)}\big[|\omega|^2\!+\!(0.2a_2)^2\big]},
\label{eq:edip-polariz1}
\\[0.0cm]
&&\beta_n(\omega)=\frac{\beta_n(0)\!-\!b_1^2\omega^2\!+\!b_2^3\,
{\rm Re}(\omega)}{(\omega^2\!-\!\omega_{\Delta}^2)^2\!+\!
|\omega^2\Gamma_{\Delta}^2|}, 
\label{eq:mdip-polariz1}
\end{eqnarray}
with $M_n$ the neutron mass, $m_\pi$ the pion mass, and 
a set of adjustable parameters given in Table~\ref{tab:pol-par}. 
The square roots in Eq.~(\ref{eq:edip-polariz1})  emulate the 
non-analytic threshold behavior related to the photoproduction of a 
pion~\cite{Lensky:2015awa,Griesshammer:2012we} and 
Eq.~(\ref{eq:mdip-polariz1}) takes the form of an energy-dependent 
Breit-Wigner that incorporates the physics of the $\Delta$ resonance. 
Our parameters are fitted to the theoretical curves of Lensky 
{\em et al.}~\cite{Lensky:2015awa} in three different ways. In Set 1 we let 
$\alpha_n(0)$ and $\beta_n(0)$ be free parameters, in Set 2 we fix them to 
the PDG central value~\cite{Patrignani:2016xqp}, and in Set 3 we fix them to 
the central value of Kossert {\em et al.}~\cite{Kossert:2002ws}.
The quality of the parametrization can be seen in Fig.~\ref{fig:n-polariz}
and is satisfactory for our purposes, falling
well within the theoretical errorbars~\cite{Lensky:2015awa}. In particular, 
on the left panel one sees the cusp behavior associated to the pion 
photoproduction, and on the right panel, the increase of $\beta_n$ near 
the delta-neutron mass difference $\sim 230$~MeV~\cite{BabbHigaHussein17}. 

\begin{table}[tbh]
\caption{Parameters of Eqs.~(\ref{eq:edip-polariz1}), (\ref{eq:mdip-polariz1})
fitted to the theoretical curves of Ref.~\cite{Lensky:2015awa}.
$\alpha_n(0)$ and $\beta_n(0)$ units are $10^{-4}{\rm fm}^3$, the remaining 
ones in MeV.
\label{tab:pol-par}}
\begin{center}\begin{tabular}{c||c|c|c|c|c|c|c|c}
& $\alpha_n(0)$ & $a_1$ & $a_2$ &$\beta_n(0)$ & $b_1$ & $b_2$ &
$\omega_{\Delta}$ & $\Gamma_{\Delta}$ \\ \hline
Set 1 & 13.9968 & 12.2648 & 1621.63 & 4.2612 & 8.33572 & 22.85 & 241.484 & 66.92
65 \\ \hline
Set 2 & 11.6 & 2.2707 & 2721.47 & 3.7 & 8.67962 & 24.2003 & 241.593 & 68.3009 \\
 \hline
Set 3 & 12.5 & 5.91153 & 2118.79 & 2.7 & 9.27719 & 26.328 & 241.821 & 70.8674 \\
\end{tabular}\end{center}
\end{table}

As shown in the following, the basic inputs to our CP interactions are 
the dynamic dipole polarizabilities $\alpha_n$ and $\beta_n$ evaluated at 
imaginary frequencies. To make sure our 
Eqs.~(\ref{eq:edip-polariz1}) and (\ref{eq:mdip-polariz1}) are reasonable in 
the complex domain we make a numeric comparison of these parametrizations 
with the heavy-baryon chiral perturbation theory (HB-$\chi$PT) expressions of 
Hildebrandt {\em et al.}, given in Appendices B and C of 
Ref.~\cite{Hildebrandt:2003fm}. The latter are given by the thin solid lines 
in Fig.~\ref{fig:n-polariz}, for real photon energies. The same expressions 
were extended to imaginary energies, and we checked that agree with our 
parametrizations up to $i\omega\lesssim i\,m_\pi$ 
(see~\cite{BabbHigaHussein17} for a detailed discussion). 

\section{Neutron under Casimir-Polder forces}

In this Section we recollect the main formulas and results from our 
previous works~\cite{BabbHigaHussein17,HBH17-2,Higa:2018yav}. We consider 
only the parameters from Set 1, which represents qualitatively the other 
sets. 

The CP interactions between two neutrons is given 
by~\cite{FeiSuc70,SprKel78a,Bab10,BabbHigaHussein17} 
\begin{eqnarray}
&&V_{CP,nn}(r)=-\frac{\alpha_0}{\pi r^6 }\,I_{nn}(r)\,,
\nonumber\\[0mm]
&&I_{nn}(r)=\int_0^{\infty}d\omega\,e^{-2\alpha_0\omega r}
\Big\{\Big[\alpha_n(i\omega)^2+\beta_n(i\omega)^2\Big]P_E(\alpha_0\omega r)
\nonumber\\[0mm]
&&\hspace{3.0cm}
+\Big[\alpha_n(i\omega)\beta_n(i\omega)
+\beta_n(i\omega)\alpha_n(i\omega)\Big]
P_M(\alpha_0\omega r)\Big\}\,,
\nonumber\\[0mm]
&&P_E(x)=x^4+2x^3+5x^2+6x+3\,,\quad P_M(x)=-(x^4+2x^3+x^2)\,,
\label{eq:Vcpnn}
\end{eqnarray}
where $\alpha_0\approx 1/137$ is the electromagnetic fine structure constant. 
Due to the exponential factor $\exp(-2\alpha_0\omega r)$ in the above formula, 
it is straightforward to check that the asymptotic region $r\to\infty$ is 
dominated by frequencies $\omega\to 0$. In the static limit, the integral 
can be performed analytically and one arrives at the original Casimir-Polder 
result, $V_{CP,nn}^{\star}(r)=V_{CP,nn}(r\to\infty)=-\big[23(\alpha_n^2(0)
+\beta_n^2(0))-14\alpha_n(0)\beta_n(0)\big]/(4\pi r^7)$. The static limit 
serves as a numerical check, though it happens at distances much larger than 
the hadronic/nuclear scale of a few fm, as we discuss in the following. 
As one moves inwards, the effects of frequency-dependent polarizabilities 
become apparent from low to high values of $\omega$. 
\begin{figure}[tbh]
\begin{tabular}{lr}
\includegraphics[width=0.57\textwidth,clip]{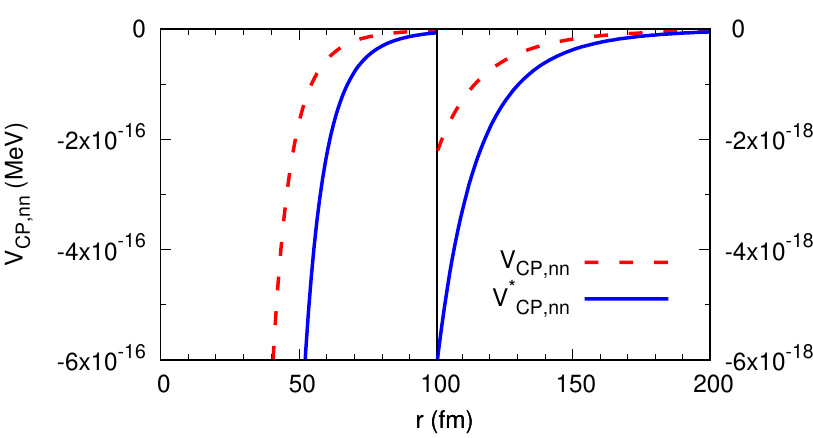}&
\includegraphics[width=0.43\textwidth,clip]{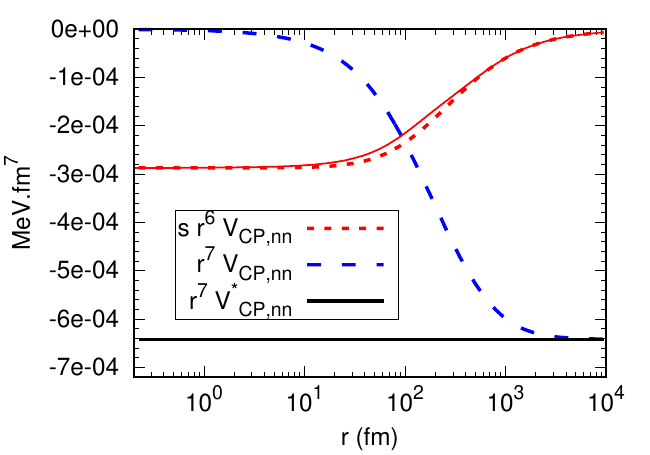}
\end{tabular}
\caption{\protect
Results for the CP-interaction between two neutrons. 
Adapted from~\cite{BabbHigaHussein17}. 
}
\label{fig:Vcp-nn}
\end{figure}

Fig.~\ref{fig:Vcp-nn} shows the behavior of the CP potential between two 
neutrons (dashed line), compared to the static limit (solid line). In the 
left panel one sees a quenching in the strength of the interaction due to 
the dependence on the frequency of the polarizabilities. The right panel 
allows one to quantify better the large distance behavior. The short-dashed 
curve is the CP potential multiplied by $s\,r^6$, where $s=100$~fm to fit 
in the figure. The long-dashed and thick-solid curves stand for the dynamic 
($V_{CP,nn}$) and static ($V_{CP,nn}^{\star}$) polarizabilities versions of 
the potential, respectively, multiplied by $r^7$. The thin solid curve 
is the arctan parametrization~\cite{OCaSuc69} that connects the $1/r^7$ 
asymptotic CP and the mid-distance $1/r^6$ vdW 
behaviors. One sees that at small distances $r\lesssim 20$~fm there is a 
clear $1/r^6$ behavior, meaning that the integrand of Eq.~(\ref{eq:Vcpnn})
is nearly constant. From the exponential factor one concludes that this 
region is probing neutron excitations larger than 
$(2\alpha_0\times 20\,{\rm fm})^{-1}\sim 670$~MeV. The $\Delta$ resonance 
has its biggest influence around $(2\alpha_0\omega_{\Delta})^{-1}\sim 50$~fm, 
though it contributes primarily to $\beta_n$, which is much smaller than the 
$\alpha_n$ contribution. The energy related to the pion production threshold 
affects distances around $(2\alpha_0\omega_{\pi})^{-1}\sim 100$~fm. The 
asymptotic $1/r^7$ behavior is achieved only beyond $10^3$~fm, due to 
dynamic polarizabilities with frequencies $\omega_\gamma\lesssim 10$~MeV. 

For the neutron-Wall (nW) CP potential one 
has~\cite{ZhouSpruch95,YanDalBab97,BabbHigaHussein17}
\begin{eqnarray}
&&V_{CP,nW}(r) = -\frac{\alpha_0}{4\pi r^3}J_{nW}(r)\,,
\quad J_{nW}(r)=\int_0^{\infty}d\omega\,e^{-2\alpha_{0}\omega r}
\alpha_{n}(i\omega)Q(\alpha_{0}\omega r)\,,
\nonumber\\[0mm]
&&Q(x)=2x^2+2x+1\,,
\label{eq:integ_nW}
\end{eqnarray}
where for this pilot study, we consider only the electric polarizability 
$\alpha_n$ component. 
(The magnetic polarizability term of the total nW CP potential enters
with the opposite sign~\cite{Boy69},
though for the neutron $\alpha_n(0)/\beta_n(0)\sim 3$,
so one might  view  Eq.~(\ref{eq:integ_nW}) 
as the most optimistic estimate of the effect.)
The asymptotic 
limit of Eq.~(\ref{eq:integ_nW})  gives $V_{CP,nW}^{\star}(r)=V_{CP,nW}(r\to\infty)
=-3\alpha_n(0)/(8\pi r^4)$. 

\begin{figure}[tbh]
\begin{tabular}{lr}
\includegraphics[width=0.57\textwidth,clip]{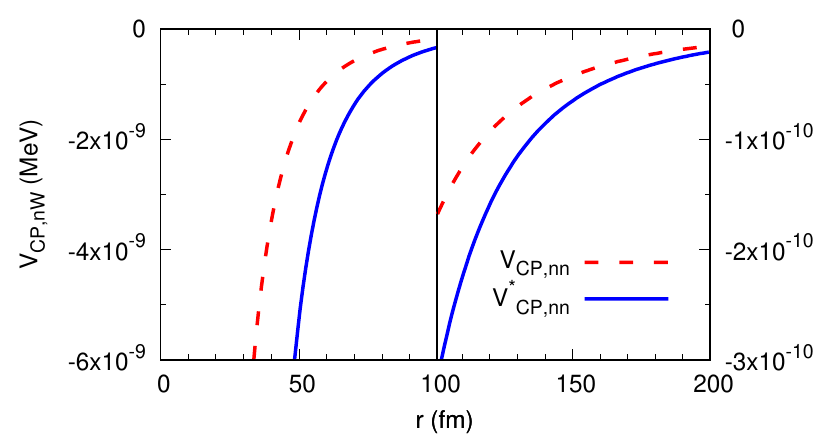}&
\includegraphics[width=0.43\textwidth,clip]{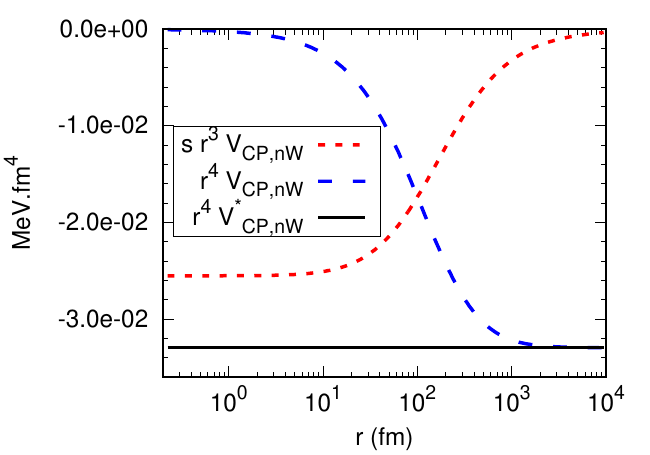}
\end{tabular}
\caption{\protect
Results for the CP-interaction between a neutron and a wall. 
Adapted from~\cite{BabbHigaHussein17}. 
}
\label{fig:Vcp-nw}
\end{figure}

Fig.~\ref{fig:Vcp-nw} shows the CP-interaction between a neutron and a wall, 
as a function of the separation $r$. On the right panel the short-dashed 
curve is multiplied by $s\,r^3$ with $s=100$~fm. The long-dashed and 
thick-solid curves are analogous to the $V_{CP,nn}$ case, multiplied by 
$r^4$ instead. The qualitative features of the mid-distance $1/r^3$ and 
the asymptotic $1/r^4$ behaviors are practically the same as the 
$V_{CP,nn}$ case. 

For two walls separated by a distance $L$ and one neutron 
in between, at a distance $z$ from the 
midpoint~\cite{ZhouSpruch95,YanDalBab97,BabbHigaHussein17}, the CP-potential 
reads
\begin{align}
&V_{CP,WnW}(z, L)=
\nonumber\\
&-\frac{1}{\alpha_{0}\pi L^4}\int_{0}^{\infty}u^3du\,
\alpha_n\left(i\frac{u}{\alpha_{0}L}\right)
\int_{1}^{\infty}\frac{dv}{\sinh(uv)}\left[
v^2\cosh\left(\frac{2z}{L}uv\right)-e^{-uv}\right]\,.
\label{eq:Vwnw01}
\end{align}
In the static limit, the integral can be done analytically and leads to 
\begin{align}
V_{CP,WnW}^{\star}(z,L) &= -\frac{\alpha_{n}(0)}{\alpha_{0}\pi L^4}\left\{
\frac{3}{8}\left[\zeta\left(4,\frac{1-f}{2}\right)
+\zeta\left(4,\frac{1-f}{2}\right)\right]-\frac{\zeta(4,1)}{4}\right\}
\nonumber\\
&= -\frac{\pi^{3}\alpha_{n}(0)}{\alpha_{0}L^4}\left[
\frac{3 - 2\cos^{2}(\pi f/ 2)}{8\cos^{4}(\pi f/ 2)}
- \frac{1}{360}\right],
\label{eq:Vwnw02}
\end{align}
where $f=2z/L$ and
\begin{equation}
\zeta(a,b)=\sum_{k=0}^{\infty}\frac{1}{(k+b)^a}
\end{equation}
is the generalized Zeta function. Eq.~(\ref{eq:Vwnw02}) explicitly shows 
the asymptotic $L^{-4}$ behavior of $V_{CP,WnW}$. 
\begin{figure}[tbh]
\begin{tabular}{lr}
\includegraphics[width=0.47\textwidth,clip]{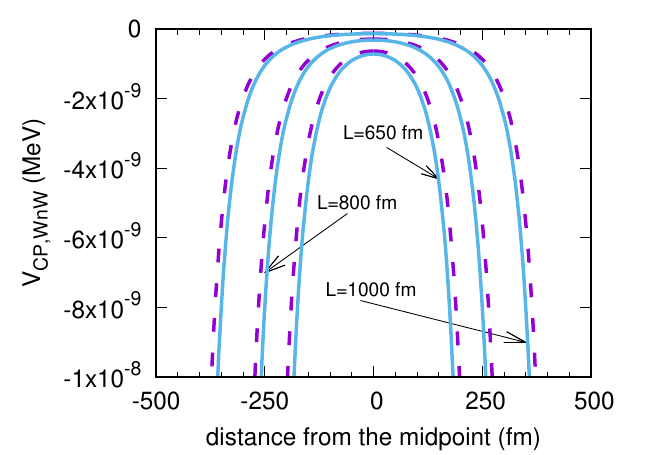}&
\includegraphics[width=0.47\textwidth,clip]{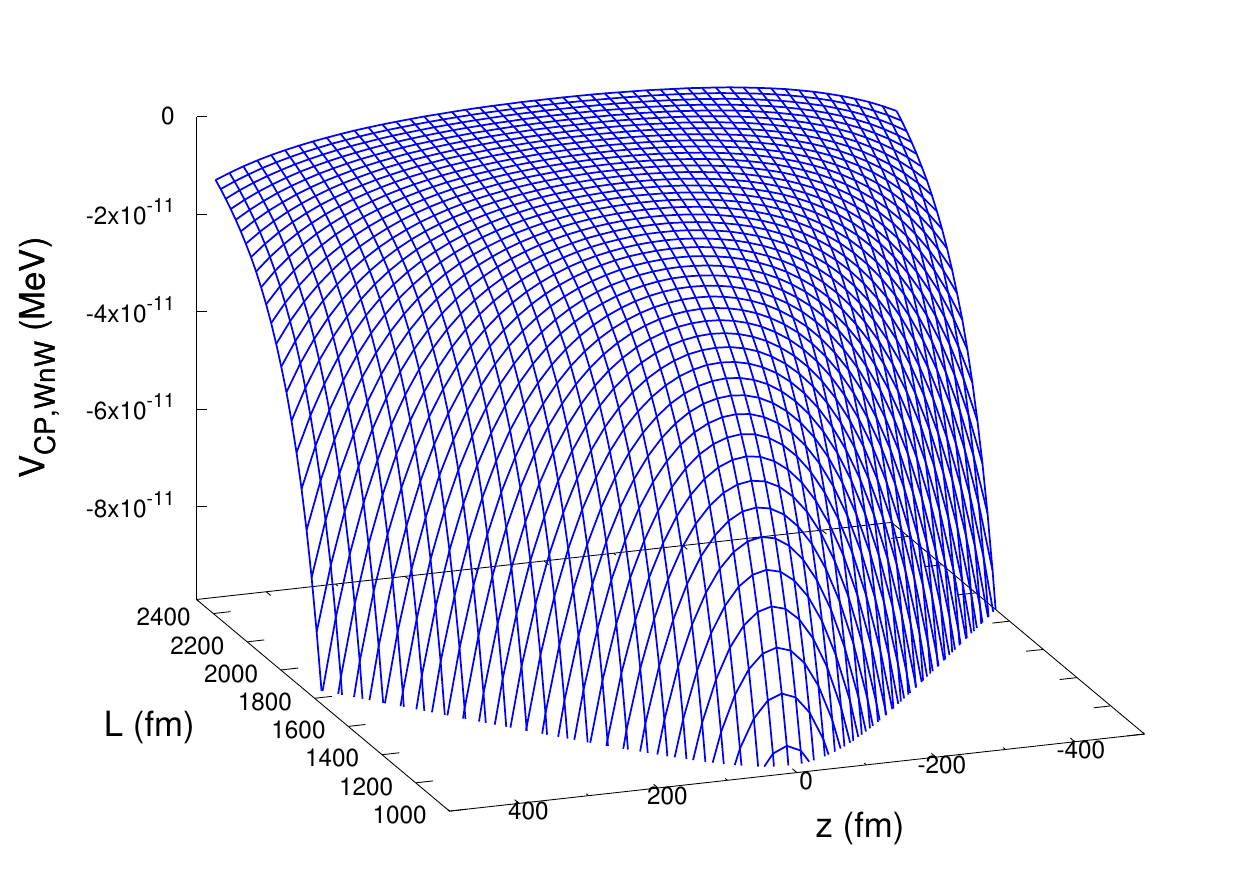}
\end{tabular}
\caption{\protect
Results for the CP-interaction of a newtron between two walls, as a function 
of the neutron distance from the midpoint $z$. 
Adapted from~\cite{BabbHigaHussein17}. 
}
\label{fig:Vcp-wnw}
\end{figure}

Fig.~\ref{fig:Vcp-wnw} shows the CP-interaction of a neutron between two 
walls. The right panel shows its dependence of $V_{CP,WnW}$ on both $L$ and 
$z$ while on the left panel one has the $z$ dependence of both 
$V_{CP,WnW}$ (dashed line) and $V_{CP,WnW}^\star$ (solid line), for three 
selected values of $L$. 

\section{Discussions and concluding remarks}

In this work, we extend to neutron physics the phenomenology of the 
CP forces developed in atomic and molecular physics. 
We present the CP-interactions between two neutrons, a neutron and a wall, 
and a neutron between two walls. This work goes beyond the static limit, 
taking into consideration the frequency-dependence of the electric and 
magnetic dipole polarizabilities of the neutron. It embraces the same spirit 
as the work by Spruch and Kelsey~\cite{SprKel78a} regarding dynamic 
polarizabilities. 

One finds that the CP-interactions between two neutrons and between a neutron 
and a wall have their long-distance behavior driven by the low-energy 
dynamics of the Compton sub-amplitude. Chiral dynamics provides reliable 
predictions for Compton scattering observables up to around the excitation 
energy of the $\Delta$ resonance, $\sim 300$~MeV. Therefore our results are 
not reliable for distances shorter than $\sim 30$~fm. The low-energy dynamics 
associated with the $\Delta$ resonance and the one-pion photoproduction 
dictate the behavior of the CP-interactions around $50$~fm and $100$~fm, 
respectively. One observes the smooth transition from the vdW-like 
to the asymptotic CP-like behavior over a range as large as $r\sim 10^3$~fm,
though 
only beyond such distances do our CP-interactions reach the expected static 
limit. 

In the asymptotic $1/r^7$ regime, the value of the neutron-neutron CP 
potential may be too small to be of any relevance to hadronic/nuclear 
physics. However, in the physics of ultracold neutrons the slower $1/r^4$ tail 
of the neutron-wall and the wall-neutron-wall CP potentials may compete with 
other important effects. For instance, the repulsive Fermi pseudo-potential 
energy close to the surface of nickel and aluminium is about $252$~neV and 
$54$~neV, respectively~\cite{BabbHigaHussein17}. This is comparable to 
the value of the neutron-wall CP interaction at $r\sim 1500$~fm. 
Clearly, a more quantitative estimate of these effects ought to be carried 
out by experiments aiming at confinement of ultracold neutrons. 
For instance, our Eqs.~(\ref{eq:integ_nW})  and (\ref{eq:Vwnw01}) take into account 
only the electric dynamic dipole polarizability. 
Contributions from the magnetic polarizability were addressed in the static 
limit in~\cite{Boy69} and are expected to be non-negligible. 
Besides, at this length scale the perfect conducting wall approximation 
used here as in atomic physics does not model effects due to the spatial 
extension of atoms and their arrangement in a real metal condition; these 
and other considerations might be especially important for neutron distances 
very close to the wall.

Other possible places where CP interactions may have some relevance are 
in systems with three and four neutrons~\cite{HBH17-2}. The existence of 
bound, virtual, or resonant states in these systems is an old and persistent 
question in few-body nuclear physics. While bound states are quite improbable 
due to constraints of the nuclear interactions fitted to other nuclei, 
the existence of three or four neutron resonances remain a controversial 
topic~\cite{PhysRevLett.117.182502,PhysRevLett.118.232501,PhysRevLett.119.032501,PhysRevC.93.044004,PhysRevLett.123.069201}, 
especially due to a recent observation of a signal compatible with a 
four-neutron resonance~\cite{PhysRevLett.116.052501}. 
Without dealing with the nuclear interaction part which can be quite 
involved, one asks if a long-range electromagnetic interaction such as the 
CP force is able to change the position or either the nature of the possible 
three and four neutron states. The potential energy due to the CP force 
was estimated in~\cite{HBH17-2} for two different configurations of three 
neutrons, and one for four neutrons. An equilateral triangle arrangement 
of neutrons with sides of length $r$ gives a repulsion of 
$\sim 1.73\,\hbar c\alpha_n^3/(\pi r^{10})$ and a linear chain of three 
neutrons equally separated by $r/2$ gives an attraction of 
$\sim -186\,\hbar c\alpha_n^3/(\pi r^{10})$. Four neutrons in a tetrahedron 
configuration with edge length $r$ leads to an attraction of 
$\sim -633\,\hbar c\alpha_n^4/(\pi r^{13})$.

Mahir was a frequent visitor to The Institute for Theoretical Atomic,
Molecular, and Optical Physics (ITAMP), gave seminars in 1995,
1996, 2000, and 2011, and  collaborated widely in active discussions with Institute
staff and postdoctoral fellows resulting in publications with
Vasili Kharchenko, Robin C\^{o}t\'{e}, Eddy Timmermans, Paolo Tommasini, and Jack Wells.

Our work on neutron and proton Casimir-Polder forces
originated during Mahir's 2011 visit to ITAMP, during
which he became aware of available \textit{frequency-dependent} electric
and magnetic dipole polarizabilities of $n$ and $p$; complementing 
the well-known \textit{static} polarizabilities. JFB learned of Mahir's  earlier
work (from 1990)  proposing~\cite{HusLimPat90}
a method to look for color (QCD) van der Waals forces~\cite{AppFis78,FeiSuc79}
that inspired 
an experiment~\cite{VilMitLep93} 
and Mahir believed that the time had come to take a fresh look at QED van der Waals forces amongst neutrons and protons.
His intuition was correct---we found only the previous related study (discussed in Sec.~\ref{sec:intro})
from 1973~\cite{Arn73}.

Mahir's 
imagination and
expertise in the quantum mechanics of atomic, nuclear, and molecular systems is
evident in his many works, collaborations, and services to science. 
His respectability and leadership in the Brazilian nuclear physics 
community is attested by the formation of generations of nuclear 
scientists, his recognized scientific production and vision, and his pivotal 
role in promoting the construction of the rare isotope beam facility RIBRAS 
at the University of S\~ao Paulo. 
Mahir was truly an ``ambassador of physics'' and he will be greatly missed.

\bibliographystyle{spphys}       
\bibliography{casamo}   

\end{document}